\documentclass[prb,showpacs,preprintnumbers,amsmath,amssymb,twocolumn]{revtex4}

\usepackage{graphicx}
\usepackage{dcolumn}
\usepackage{bm}
\usepackage{color}

\begin{document}


\title{Relaxation and Readout Visibility \\ of a Singlet-Triplet Qubit in an Overhauser Field Gradient}


\author{C.~Barthel$^{1,*}$}
\author{J.~Medford$^{1,*}$}
\author{H.~Bluhm$^{1,\dagger}$}
\author{A.~Yacoby$^{1}$}
\author{C.~M.~Marcus$^1$}
\author{M.~P.~Hanson$^2$}
\author{A.~C.~Gossard$^2$}

\affiliation{$^1$Department of Physics, Harvard University, Cambridge, Massachusetts 02138, USA\\
$^2$Materials Department, University of California, Santa Barbara, California 93106, USA\\
}

\begin{abstract}
 Using single-shot charge detection  in a GaAs double quantum dot, we investigate spin relaxation time ($T_1$) and readout visibility of a two-electron singlet-triplet qubit following single-electron dynamic nuclear polarization (DNP). For magnetic fields up to 2 T, the DNP cycle is in all cases found to increase Overhauser field gradients, which in turn decrease $T_1$ and consequently reduce readout visibility. This effect was previously attributed to a suppression of singlet-triplet dephasing under a similar DNP cycle.  A model describing relaxation after singlet-triplet mixing agrees well with experiment. Effects of pulse bandwidth on visibility are also investigated.

\end{abstract}


\maketitle
\section{Introduction}
Confined few-electron systems are of interest for exploring spin coherence and controlled entanglement,~\cite{Chirolli} as probes of mesoscopic nuclear spin environments,~\cite{deSousa,ReillyCorr07} and as qubits for quantum information processing.~\cite{Klauser,Hanson2007} The singlet-triplet basis of two electron spins in a double quantum dot has been widely investigated as a qubit with immunity to dephasing from fluctuating uniform magnetic fields.~\cite{Levy02} An important source of both spin dephasing and relaxation in GaAs devices is hyperfine coupling to nuclear spins in the host material. The slow evolution of longitudinal Overhauser fields allows echo techniques to recover phase coherence~\cite{petta05,bluhm10,barthel10}, while even static gradients of Overhauser fields can induce triplet-to-singlet relaxation,~\cite{JohnsonT1} limiting readout fidelity.~\cite{Singleshotpaper} It is therefore important to understand how gradients in local Zeeman fields, either from micromagnets~\cite{PioroLadriere, Obata} or Overhauser fields,~\cite{Foletti09} affect singlet-triplet qubit relaxation, particularly during readout.

Dynamic nuclear polarization (DNP) using cyclic single-spin transitions can transfer angular momentum from electrons in the double dot (refreshed from reservoirs) into the host nuclear system, inducing a net nuclear polarization.~\cite{Petta08,ReillyDiff08,Foletti09}  In Ref.~\onlinecite{ReillyZamboni}, it was observed that for tens of seconds following the application of the DNP cycle, the probability, $P_S$, to measure a singlet outcome, after allowing a prepared singlet to evolve in separate dots, remained close to unity. This surprising observation was interpreted as the DNP cycle having reduced the difference in Overhauser fields between the two dots below the normal (thermal) fluctuation level while inducing a net polarization. That interpretation was consistent with some theoretical results,~\cite{HuZamboni,BurkhardZamboni,StopaZamboni} but at odds with subsequent experiment \cite{Foletti09} and more recent theory.~\cite{GullandZamboni}

In this paper, we show that over a broad range of applied magnetic fields, the DNP pumping cycle investigated in Refs.~\onlinecite{Petta08,ReillyDiff08,ReillyZamboni,Foletti09} enhances rather than reduces the gradient in nuclear polarization, along with inducing an average polarization.  
Rapidly repeated single-shot readout~\cite{Singleshotpaper} reveals that the enhanced nuclear gradient leads to a reduction in the visibility of qubit precession. We investigate qubit readout visibility as a function of nuclear field gradient, applied magnetic field, and gate voltage configuration during the measurement step of a cyclic pulse sequence. Simultaneously, triplet relaxation at the measurement point is measured in the time domain, taking advantage of fast readout electronics. We find that the dominant reduction in visibility for large nuclear polarizations is due to increased relaxation of the $m=0$ triplet during measurement, independent of applied magnetic field. We develop a model describing triplet decay via charge relaxation after rapid singlet-triplet mixing driven by a Zeeman field difference between dots, including effects of finite pulse bandwidth. The model is found to be in very good agreement with experimental results.  These results suggest an alternative interpretation of the increased singlet measurement probability following DNP,\cite{ReillyZamboni} which is that the enhanced nuclear field gradient induced by DNP causes rapid relaxation of the triplet state during measurement, which in turn diminishes  measurement  visibility while the nuclei are out of equilibrium.

\begin{figure}
\includegraphics{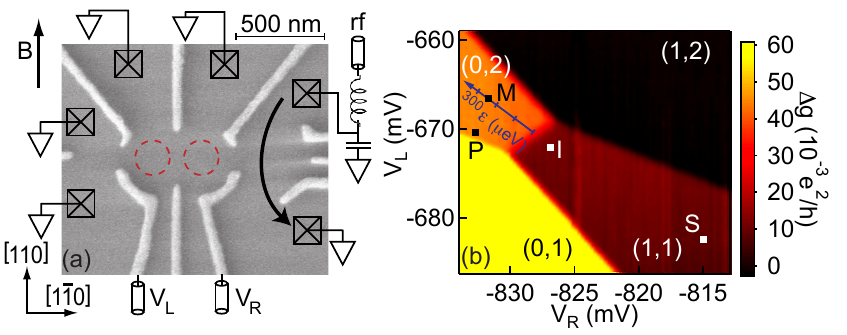}
\caption{\label{technical} (a) (Color online) (a) Micrograph of device lithographically identical to the one measured. Gate voltages, $V_{\rm{L}}$ and $V_{\rm{R}}$, set the electrostatic energy of left and right dot. A sensor quantum dot on the right allows fast measurement of the double dot charge state via rf reflectometry. The direction of applied magnetic field, $B$, is indicated, as well as the GaAs crystal axes.
(b) Change of sensor dc conductance $\Delta g$, with double dot charge state $(N_{\rm{L}},N_{\rm{R}})$, constrained to (1,1), (0,2) in this work. The qubit state is controlled by the (1,1)-(0,2) energy detuning, $\epsilon$, set by gate voltages $V_{\rm{L}}$ and $V_{\rm{R}}$ along the diagonal axis through the markers S, M. The scaling of detuning is $\left|\epsilon\right|=\eta\sqrt{\Delta V_{\rm{L}}^2 + \Delta V_{\rm{R}}^2}$, with a lever arm $\eta=40~\mu{\rm{eV/mV}}$ and voltage detunings, $\Delta V_{\rm{L}}$, $\Delta V_{\rm{R}}$, from the (1,1)-(0,2) charge degeneracy.~\cite{leverarmdisclaim}  Markers indicate gate voltages during pump- and probe-cycles. Singlet preparation at point P. Pump: $S$-$T_+$ mixing at point I, see text. Probe: Separation of singlet for $S - T_0$ mixing at point S and measurement at point M at variable detuning, $80~\mu\rm{eV}<\epsilon_{\rm{M}}<260~\mu\rm{eV}$. 
}
\end{figure}

The remainder of the paper is organized as follows. Section~\ref{system} describes the double dot system and the experimental setup. The theory of the two-electron qubit system and nuclear pumping is briefly reviewed in the first part of section~\ref{theory}. The second part of section~\ref{theory} discusses mechanisms of spin relaxation during measurement and introduces a model of these effects. Experimental results are presented in section~\ref{experiment}, beginning with the measurement of nuclear gradients and precession visibilities. Observed connections between visibility, relaxation time and Overhauser field difference are then presented, along with data showing the influence of limited pulse bandwidth. A summary of results and conclusions are given in section~\ref{conclusions}.
\section{System}
\label{system}
Double quantum dots were formed by Ti/Au depletion gates on a GaAs/Al$_{0.3}$Ga$_{0.7}$As heterostructure with a two-dimensional electron gas (2DEG) of density $2\times10^{15} ~\rm{m}^{-2}$ and mobility $20~\rm{m}^2$/Vs, 100 nm below the wafer surface. Except where noted, a field of 200 mT was applied in the direction shown in Fig.~1(a) using a vector magnet. Measurements were performed in six cool-downs of four devices for magnetic fields along all three crystal axes [Fig.~\ref{technical}(a)] over a range of applied magnetic fields from $10$ mT to $2$ T. The same overall phenomenology was found in all measurements. Results are reported here for one of those devices.

A proximal radio-frequency sensor quantum dot (SQD)  [Fig.~1(a)] was used to sense the charge state of the double dot.\cite{Reillyapl07,Dotsensorpaper} Reflectometry measurement on the SQD provides an output voltage, $v_{{\rm{rf}}}$, with good signal-to-noise on  sub-$\mu$s time scales.\cite{Dotsensorpaper}  The SQD was energized only during readout to avoid disturbance during gate operations.  Gate voltages $V_{\rm{L}}$ and $V_{\rm{R}}$, pulsed using a Tektronix AWG5014, controlled charge occupancies  $N_{\rm{L}}$ and $N_{\rm{R}}$ of the left and right dots. The charge state ($N_{\rm{L}}, ~N_{\rm{R}}$) was restricted to (1,1) and (0,2), and was controlled by gate voltages along an axis of energy detuning, $\epsilon$, running between separation  (S) and measurement (M) points [Fig.~1(b)]. Detuning  scales as $\left|\epsilon\right|=\eta\sqrt{\Delta V_{\rm{L}}^2 + \Delta V_{\rm{R}}^2}$, where $\Delta V_{\rm{L}}$ and $\Delta V_{\rm{R}}$ are gate voltages relative to the charge transition point, and $\eta=40~\mu{\rm{eV/mV}}$ is the voltage-to-energy lever arm, calibrated via transport through the double dot.\cite{JohnsonSpinBlockade,leverarmdisclaim,VanderWiel} Note that the two gates contribute symmetrically to detuning, as observed experimentally. The influence of  $V_{\rm{L}}$ and $V_{\rm{R}}$ on the interdot tunnel coupling is found to be small for the range of voltages used, and is neglected in the model presented below. 

\section{Model}
\label{theory}
\begin{figure}[t]
\includegraphics{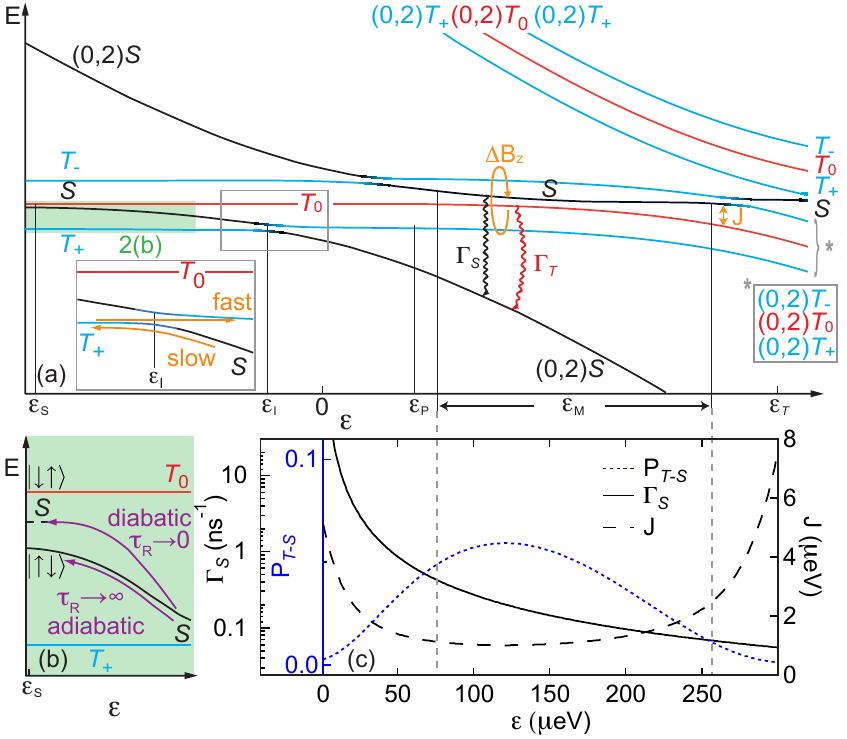}
\caption{\label{levels} (a) (Color online) (a) Energy level diagram as function of detuning, $\epsilon$, for the (1,1) charge state except where noted. Pulse-cycle detunings, $\epsilon_{\rm{P}}$ of singlet preparation, $\epsilon_{\rm{I}}$ of $S$-$T_+$ resonance, and $\epsilon_{\rm{S}}$ of $S$-$T_0$ precession, are labeled. The relaxation channels of triplet state, $T_0$, during measurement at $\epsilon_{\rm{M}}$ are indicated:  Charge relaxation, $\Gamma_S$, after $S$-$T_0$ mixing by nuclear field difference,  $\Delta B_z$, and processes not involving $\Delta B_z$, at rate $\Gamma_{T}$. The $S$-$T_0$ mixing is suppressed by the exchange energy splitting, $J$, due to two anticrossings, at $\epsilon=0$ between singlet states of (1,1)-(0,2), and at $\epsilon=\epsilon_T\sim300~\mu\rm{eV}$ between triplet states of (1,1)-(0,2). Inset: Pump cycle around $S$-$T_+$ anticrossing at $\epsilon_{\rm I}$. (b)  Ramping  detuning to  $\epsilon_{\rm{S}}$ over a time $\tau_{\rm{R}}$ converts an initial $S$ into an admixture of $S$ and $\mid\uparrow\downarrow\rangle$ with amplitudes that depend on $\tau_{\rm{R}}$. (c) Charge relaxation rate, $\Gamma_S$, of metastable $S$ state [Eq.~(\ref{GammaSeqn})]; singlet probability, $P_{T{\text-}S}$, from admixture [Eq.~(\ref{STmixeqn})]; exchange $J$  [Eq.~(\ref{Jofepseqn})], as functions of detuning, $\epsilon$ using experimental parameters (see text).
}
\end{figure}

The dependence of the two-electron energy levels on detuning,  $\epsilon$, is shown in Fig.~2(a) for the regime relevant to singlet-triplet qubit operation. The two-level system that forms the qubit is the two-electron singlet, $S$, and the $m=0$ triplet, $T_0$, of the (1,1) charge state. Preparation of the $S$ state is achieved through relaxation into the (0,2) singlet state via electron exchange with the leads at point P [see Figs.~\ref{technical}(b) and~\ref{levels}(a)]. The (0,2) singlet can be separated into the (1,1) singlet, $S$, ($+z$ on the qubit Bloch sphere) by following the lower branch of the singlet anticrossing through $\epsilon=0$ [anticrossing of black curves in Fig.~2(a)].

A net polarization of nuclei in the host GaAs substrate in the vicinity of the double quantum dot can be created electrically by cycling $\epsilon$ through the anticrossing of the singlet $S$ and the $m=1$ triplet, $T_+$, at point I~[inset of Fig.~\ref{levels}(a)].\cite{Petta08,Foletti09,ReillyDiff08,ReillyZamboni}
First, moving slowly through the anticrossing, an electron spin is flipped and a nuclear spin is flopped via hyperfine interaction. The system is then quickly brought to $\epsilon > 0$, without spin flip, and is reset to a singlet state at P via electron exchange with the leads. Ideally, the nuclear pumping cycle flips one nuclear spin per cycle but in practice the efficiency is typically lower.\cite{Brataas11}  

In an applied magnetic field, ${\bf B}$, whose direction defines the $z$ direction in real space, qubit states at the separation point, S, 
are split by the {\em difference} in the $z$ components of Zeeman fields (including nuclear Overhauser fields), $\Delta B_{\rm z}$, between left and right dots. This causes a precession between $S$ and $T_0$ at  frequency
\begin{equation}\label{Precessionfreq}
f_{\rm{S}} = \frac{\left| g \right| \mu_{\rm{B}} \Delta B_z}{h},
\end{equation}
 where $h$ is Planck's constant, $\mu_{\rm{B}}$ is the Bohr magneton, and $g\sim -0.4$ is the electron g-factor in GaAs.  A frequency shift away from $f_{\rm{S}}$ due to residual exchange at the separation point, $J_{\rm{S}}\sim10~{\rm{neV}}(\sim0.5$~mT), is usually much smaller than $g\mu_{\rm{B}}\Delta B_z$ in the present experiments. A nonvanishing $J_{\rm{S}}$ also reduces readout visibility,
 \begin{equation}
 \label{CL}
 V_J = \frac{\Delta B_z^2} {\Delta B_z^2 + (J_{\rm{S}}/g^*\mu_{\rm{B}})^2}< 1,
 \end{equation}
as previously shown theoretically \cite{coish05} and experimentally.~\cite{Laird06} In this work, however, $V_J\sim1$; the reduced visibility  arises from other sources, as discussed below. 

The sensitivity of qubit relaxation and readout visibility to gradients in Overhauser fields was investigated using a probe cycle following a sequence of DNP cycles.~\cite{ReillyCorr07,Singleshotpaper,Foletti09} The probe cycle  prepares a spin singlet in $(0,2)$, separates to point S for a time $\tau_{\rm{S}}$, then returns to the measurement point M. If the two electrons are in a singlet configuration, they return to $(0,2)$; if they are in a triplet configuration, they remain in $(1,1)$. Superpositions are projected to one of the two charge states during measurement. Spin-to-charge conversion requires the measurement time be shorter than the relaxation time of the metastable triplet to the (0,2) singlet. 

The probability of measuring a singlet or a triplet was determined by  accumulating statistics of multiple single-shot measurements.\cite{Singleshotpaper} When the single-shot integration time $\tau_M$ was much shorter than  $T_1$, the distribution of outcomes formed two separated, noise-broadened gaussians centered at the amplitudes corresponding to singlet $(v_{\rm{rf}}^S)$ and triplet $(v_{\rm{rf}}^T)$ states. Measurement visibility can be expressed as $V_{\rm M} = F_S + F_T(T_1)-1$, where $F_S$ ($F_T$) is the singlet (triplet) fidelity, corresponding to the probability that a singlet (triplet) is correctly identified as such.\cite{Singleshotpaper} Depending on the ratio $\tau_{\rm M}/T_1$, the metastable triplet may decay into a singlet during the measurement, leading to overcounting of singlets and undercounting of triplets in the output distribution.~\cite{Singleshotpaper} Specifically---{\em and this is a key point of our analysis}---for fixed measurement time, $\tau_{\rm M}$, the measurement visibility $V_{\rm M}$ decreases with decreasing $T_1$. We note that the same reduction in visibility will be observed for time averaged readout---as opposed to single-shot readout---as was used in Refs.~\onlinecite{petta05,ReillyZamboni}.

To measure relaxation of triplet states into singlets during measurement, $v_{\rm{rf}}(t)$ was measured with high temporal resolution after moving to point M, then subsequently averaged over many successive pulse cycles. For short measurement times, $t \ll T_1$, the signal corresponds to an equal mix of singlet and triplet states, $\langle v_{\rm{rf}}(t)\rangle\sim (v_{\rm{rf}}^S+ v_{\rm{rf}}^T)/2$, while for $t \gg T_1$ the signal corresponds to the (0,2) charge state and therefore the singlet, $\langle v_{\rm{rf}}(t)\rangle\sim v_{\rm{rf}}^S$. Experimentally, we find that $\langle v_{\rm{rf}} (t)\rangle$ is approximately exponential in $t$, giving a measure of the triplet relaxation time $T_1$ at the measurement point. 

Relaxation pathways of the $m=0$ triplet at M are shown in the energy diagram in Fig.~2(a). A difference in Zeeman fields, $\Delta B_z$, between the two dots will cause rapid precession between $T_0$ and the (1, 1) singlet, $S$, which can then rapidly relax to the (0,2) singlet with a rate $\Gamma_S$ via spin-conserving phonon emission. Direct relaxation of the (1,1) triplet, at a rate $\Gamma_T$, involves a change in total spin---mediated, for instance, by electron exchange with the leads---and so is a much slower process.  

In previous measurements of triplet relaxation, exchange, $J$, at point M [see Fig.~2(a)] was intentionally set to be much smaller than $\Delta B_z$.~\cite{JohnsonT1}  In that case, a $T_0$ state brought to M would oscillate between $S$ and $T_0$ rapidly, giving an average singlet occupation of 1/2, and a decay rate $\Gamma_S/2$, independent of $\Delta B_z$. In the present measurement as well as in previous $T_2^{*}$-type experiments \cite{petta05, ReillyZamboni, Foletti09}, tunnel coupling is much larger, so that $J$ at point M is not necessarily small compared to $\Delta B_z$. The effect of sizable $J$ at the measurement point is a suppression of mixing between $T_0$ and $S$ by an amount that depends on the ratio $\Delta B_z/J$. In this case, the average $S$ occupation at M, and thus triplet decay via fast spin-conserving processes, will increase with increasing $\Delta B_z$.

Triplet decay is modeled by extending Ref.~\onlinecite{TaylorPRB2007} to include nonzero $J$. Populations of the eigenstates of the Hamiltonian $\mathcal{H} = J(\epsilon_{\rm{M}})(\sigma_z+\mathbb{I})/2 - \Delta B_z\sigma_x/2$, decay with rates $\Gamma^{\pm}=\Gamma_S(\epsilon_{\rm{M}}) |\langle S|E^{\pm}\rangle |^2 + \Gamma_T |\langle T_0|E^{\pm}\rangle |^2$. 

The eigenstates of $\mathcal{H}$ are given by
\begin{equation}
	|E^{\pm} \rangle= \frac{\Delta B_z |S\rangle + \Omega^{\pm} |T_0\rangle}{\sqrt{\Delta B_z^2 +  (\Omega^{\pm})^2}}, 
\end{equation}
where  $\Omega^{\pm} = J(\epsilon_{\rm{M}}) \pm \sqrt{ J^{2}(\epsilon_{\rm{M}})+\Delta B_z^2}$.

In principle, this results in a bi-exponential decay of the triplet probability, $P_T(t) =  P_T(0)(|\langle T_0 | E^+\rangle |^2 e^{-t \Gamma^+} +  |\langle T_0 | E^-\rangle |^2 e^{-t \Gamma^-})$, but in practice, we expect only a single exponential. This is due to the fact that for $J\gg\Delta B_z$,  $|E^-\rangle$ has a large overlap with the singlet, leading to a much larger $\Gamma^-$ and a correspondingly small overlap with $T_0$. For instance, for  large nuclear gradients, $\Delta B_z\!\sim\!35$~mT, and small exchange splittings, $J\!\sim\!1~\mu$eV, $|E^-\rangle$ accounts for roughly one eighth of the initial triplet, and decays seven times more rapidly than the triplet-like eigenstate, $|E^+\rangle$. Under such conditions, we can model $P_T(t)$ as a single exponential, $P_T(0) e^{-t/T_1}$, where \begin{equation}\label{T1eqn}
	T_1 = (\Gamma^+)^{-1}\cong[\Gamma_S(\epsilon_{\rm{M}}) P_{T{\text-}S} + (1-P_{T{\text-}S})\Gamma_T]^{-1},
\end{equation}
 and $P_{T{\text-}S}$ is the fraction of the remaining triplet that overlaps with the (1,1) singlet, 
  \begin{equation}\label{STmixeqn}
  P_{T{\text-}S} = |\langle S|E^+\rangle|^2= \frac{1}{2}\left(1-\frac{J(\epsilon_{\rm{M}})}{\sqrt{\Delta B_z^2+J(\epsilon_{\rm{M}})^2}}\right)\,
  .
  \end{equation} 
In this model, $\Gamma_T$ is governed by a decay channel that is independent of $\Delta B_z$, such as exchange with the leads.  In principle, non-spin-conserving processes that generate triplet relaxation also contribute to $\Gamma_S$, but since $\Gamma_T$ is at least two orders of magnitude smaller than $\Gamma_S$ for all $\epsilon_{\rm{M}}$ in the experiment, these contributions can be neglected. To simplify the modeling further, we assume $\Gamma_T$ does not depend on $\epsilon_{\rm{M}}$.

Exchange splitting  at the measurement point, $J(\epsilon_{\rm M})$, is affected by two charge-state anticrossings, one of the (0,2) and (1,1) singlet states, centered at $\epsilon = 0$, and the other of the (0,2) and (1,1) $T_0$ states, centered at $\epsilon = \epsilon_{T}$ [see Fig.~2(a)]. For $0 < \epsilon_{\rm{M}} <  \epsilon_T$,  the (1,1) singlet is the upper branch (denoted $\cup$) of the singlet anticrossing, and the (1,1) triplet is the lower branch (denoted $\cap$) of the $T_{0}$ anticrossing. Also, in this range the upper branch of the singlet anticrossing remains above the lower branch of the  $T_{0}$ anticrossing. We write

\begin{equation}\label{Jofepseqn}
  	J(\epsilon_{\rm M})=  E_S(\epsilon_{\rm{M}}) - E_{T_0}(\epsilon_{\rm{M}})
 \end{equation} 
using the forms given in Eqs.\,(16,17) in Ref.~\onlinecite{TaylorPRB2007}.\cite{JakeErr} The singlet energy is
  \begin{equation}
  	E_S(\epsilon) =  E_{\cup}(\epsilon) = \frac{t_S^2}{\sqrt{4\,t_S^2 + \epsilon^2}+\epsilon},
 \end{equation} 
where $t_{S}$ is the tunnel coupling for the singlet anticrossing centered at $\epsilon =0$, and
the triplet energy is 
\begin{equation}  
E_{T_0}(\epsilon) = E_{\cap}(\epsilon- \epsilon_{T}) = \frac{-t_T^2}{\sqrt{4\,t_T^2 + (\epsilon-\epsilon_T)^2}-(\epsilon-\epsilon_T)},
\end{equation}  
where  $t_{T}$ is the tunnel coupling for the $T_0$ anticrossing centered at $\epsilon = \epsilon_{T}$. 

In previous experiments, $\Gamma_S$ was found to decrease with increasing detuning with a dependence falling between  $\epsilon^{-1}$ and $\epsilon^{-2}$, consistent with expected phonon mechanisms.\cite{Fujisawa98} Specifically, piezoelectric interaction with 3D (2D) phonons gives $\Gamma_S \propto \epsilon^{-1}$ ($\epsilon^{-2}$).\cite{Fujisawa98} Here, we assume a form
\begin{equation}
\label{GammaSeqn}
\Gamma_S = \alpha \epsilon^{-1} + \beta \epsilon^{-2},
\end{equation} 
with $\alpha$ and $\beta$ fit parameters. Figure \ref{levels}(c) shows the charge relaxation rate, $\Gamma_S$ [Eq.~(\ref{GammaSeqn})], exchange, $J$  [Eq.~(\ref{Jofepseqn})], and singlet admixture, $P_{T{\text-}S}$ [Eq.~(\ref{STmixeqn})], as functions of detuning, using experimental parameters determined by fits described below.

\section{Experiment}
\label{experiment}
\begin{figure}[t]
\includegraphics{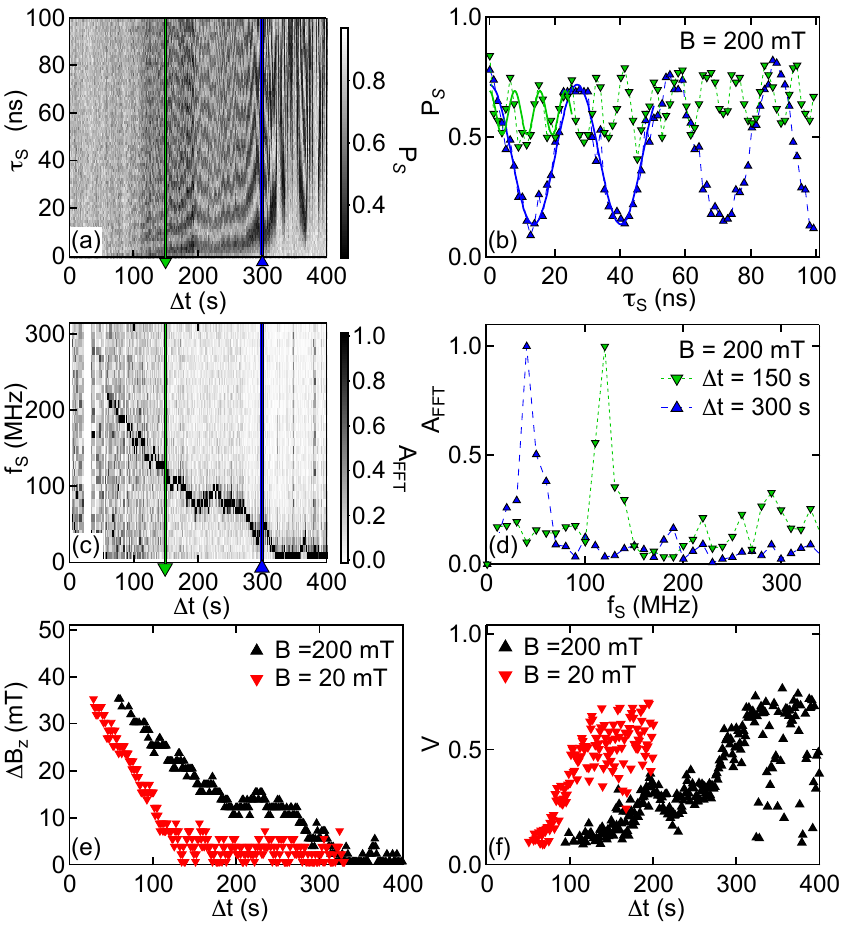}
\caption{\label{timeev}(Color online) (a) Probability, $P_S$, of singlet measurement outcome (grayscale) as function of $S$-$T_0$ mixing time, $\tau_{\rm{S}}$,  and time, $\Delta t$, after a 60 s, ${\sim4}$~MHz pump cycle; taken at $B=200$~mT. Note the near-unity singlet measurement probability with low visibility, high frequency oscillations at small $\Delta t$. (b) Vertical cuts through (a), showing $P_S(\tau_{\rm{S}})$ curves, from which visibilities, $V$, and nuclear field differences, $\Delta B_z$, are extracted via fits (solid curves) to Eq.~(\ref{PSts}), for $\Delta t = 150$~s and 300 s. (c) Normalized Fourier amplitudes, $A_{\rm{FFT}}$ (grayscale), of data in (a). (d) Vertical cuts through (c), for $\Delta t =150$~s and 300~s following pumping cycle.  (e) Overhauser field difference, $\Delta B_z$~ at $B_{\rm{ext}}=100$~mT (black) and 20 mT (red) as function of time, $\Delta t$ after pumping, using Eq.~\ref{Precessionfreq}, which gives $\Delta B_z   \sim f_{\rm{S}}~ {\rm{mT}}/(\rm{6.2~MHz})$. $\Delta B_z$ decays faster at lower fields, as expected for spin diffusion.~\cite{ReillyDiff08,ReillyCorr07} (f) Visibility, $V$, of $S$-$T_0$ oscillations as function of $\Delta t$ for data in (a) (black), and similar data at $B_{\rm{ext}}=20$~mT (red). }
\end{figure}
A 60 s pump-cycle sequence with a ${\sim4}$~MHz repetition rate, ramping through ${\sim10}~\mu$eV in $\epsilon$ around $\epsilon_I$ in 100 ns,  was used to prepare a nonequilibrium nuclear spin configuration in the double quantum dot. Immediately after, a probe-cycle sequence was run to extract the nuclear field difference from the $S$-$T_0$ precession rates. Following  Ref.~\onlinecite{Singleshotpaper}, within each cycle a singlet was prepared in (0,2) at $\epsilon_{\rm P}$, separated to $\epsilon_{\rm{S}}\sim-700~\mu$eV for a  time $\tau_{\rm{S}}$, then moved to $\epsilon_{\rm M}$ for a measurement time $\tau_{\rm{M}}^{\rm{max}}\sim10~\mu$s. The charge sensor signal, $v_{\rm{rf}}$, was integrated over $\sim300$~ns, yielding a single-shot measurement, which was identified as either a singlet or a triplet by comparison to a threshold voltage. For each $\tau_{\rm{S}}$, 100 single-shot measurements were performed, with the fraction of singlet outcomes determining $P_S$, the singlet probability.  Figure~\ref{timeev}(a) shows $P_S$ as function of $\tau_{\rm{S}}$ and time, $\Delta t$, after the pump-cycle sequence. The separation time, $\tau_{\rm{S}}$, is stepped from 1 ns to 100 ns in 80 steps. Resulting sets of 8000 cycles, acquired in 1 s [one column in Fig.~\ref{timeev}(a)], are shown for two values of $\Delta t$ in Fig.~\ref{timeev}(b). Measurements taken soon after the pump-cycle sequence, $\Delta t \lesssim 50$~s, show $P_S \sim 1$. At somewhat longer times, $50$~s $\lesssim \Delta t \lesssim 300$~s, high frequency oscillations can be seen with low visibility. At longer times, $\Delta t \gtrsim300$~s, near-unity oscillations are observed, reflecting  $S$-$T_0$ precession with frequencies corresponding to equilibrium nuclear field differences, $\Delta B_z$. 
 
Fast Fourier transform (FFT) power spectra of $P_S(\tau_{\rm{S}})$ are shown in Fig.~\ref{timeev}(c). Each spectrum has a single, strong peak, as seen in the cuts at $\Delta t=150$~s and $\Delta t=300$~s  in Fig.~\ref{timeev}(d). The peak frequency decreases with increasing $\Delta t$, indicating that the Overhauser field  gradient, $\Delta B_z$ is decreasing following the pump cycle. 

Fits of the form
\begin{equation}
\label{PSts}
P_S(\tau_{\rm{S}}) = P_0 + 1/2\,V \cos{(2 \pi f_{\rm{S}}\tau_{\rm{S}})}
\end{equation}
to the time-domain data [Fig.~3(a)] are used to extract values for $\Delta B_{z}$ using Eq.~(\ref{Precessionfreq}) and visibility, $V$. Examples of these fits are shown as solid curves in Fig.~3b.  Values  for $\Delta B_{z}$ obtained from time-domain fits are consistent with values of FFT peak frequencies [Fig.~3(c)].  For the cut at $\Delta t\sim150$~s the visibility is $V=0.2$, and the extracted nuclear field difference is $\Delta B_z \sim 20$~mT, while for the cut at $\Delta t\sim300$~s, $V=0.6$, and $\Delta B_z \sim 6$~mT.~\cite{Param3b} Figures~\ref{timeev}(e,f) show $\Delta B_z$ and $V$ for the data sets in Fig.~3(a) (time domain) and Fig.~3(c) (frequency domain).  Results of a similar pump-probe experiment at lower applied field, $B=20$~mT is also shown in Figs.~3(e,f). Note that the decay of the field difference with time, $\Delta t$, is faster for the 20~mT data, consistent with a nuclear spin diffusion model.~\cite{ReillyCorr07,ReillyDiff08}

\begin{figure}[h!]
\includegraphics{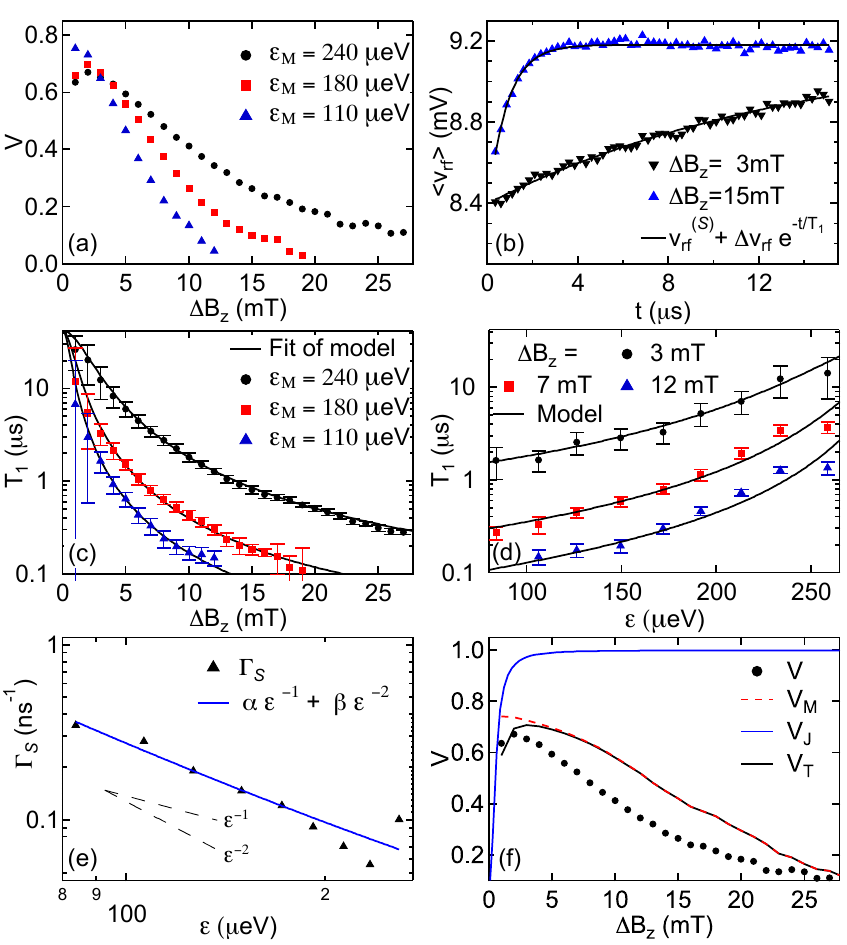}
\caption{\label{T1Fig} (Color online)    
 (a)  Parametric plot of extracted visibility, $V$, of $S$-$T_0$ precession and  nuclear field gradient, $\Delta B_z$, for three values of measurement-point detuning, $\epsilon_{\rm{M}}$. (b) rf voltage amplitude, $
\langle v_{\rm{rf}}\rangle$, as function of time at $\epsilon_{\rm{M}}$ following separation, for two values of $\Delta B_z$, averaged over an ensemble of 8000 point with separation times $\tau_{\rm{S}}$ ranging from 1 - 100 ns.  Exponential fits with offsets (solid curves, formula in graph) give $T_1\sim 13~\mu$s for $\Delta B_z= 3$~mT and $T_1\sim 0.8~\mu$s for $\Delta B_z=15$~mT.
(c) Parametric plot of relaxation time, $T_1$, of $T_0$ state versus  Overhauser gradient, $\Delta B_z$. Solid lines are a best fit of Eq.~(\ref{T1eqn}) over all values of $\epsilon_{\rm{M}}$, with a single set of fit parameters $\Gamma_S(\epsilon_{\rm{M}})$ [see (e)], $\Gamma_T\sim (40\mu{\rm{s}})^{-1}$, and tunnel coupling, $t_T=12~\mu$eV, which determines $J(\epsilon_{\rm{M}})$ via Eq.~(\ref{Jofepseqn}). (d)  $T_1$ as function of detuning $\epsilon_{\rm{M}}$, for 3 different $\Delta B_z$. Model (curves) based on Eq.~(\ref{GammaSeqn}) for $\Gamma_S(\epsilon_{\rm{M}})$ with parameters from (e). 
 (e) Singlet relaxation rate $\Gamma_S$ from fit of Eq.~(\ref{T1eqn}) to data in (c,d) along with fit to Eq.~(\ref{GammaSeqn}) (solid line) with fit parameters $\alpha\sim11~\mu{\rm{eV}}\,{\rm{ns}}^{-1}$ and $\beta \sim 1600~(\mu{\rm{eV}})^2\,{\rm{ns}}^{-1}$. The functional form is consistent with rate contributions from 3D ($\alpha$) and 2D ($\beta$) piezo-electric phonons, see Ref.~\onlinecite{Fujisawa98}.
(f) Visibility, $V$, from fits to $S$-$T_0$ precession data [Fig.~3(b)], for $\epsilon_{\rm{M}}=240~\mu$eV, along with model visibilities, $V_{\rm{T}}$ based on Eq.~(\ref{visibility}). Single-shot measurement visibility, $V_{\rm{M}}$, is calculated from the measured $T_1$ and measurement signal-to-noise ratio. The pure singlet precession visibility, $V_J$, Eq.~(\ref{CL}), reflects finite exchange, $J_{\rm{S}}\sim 10~\rm{neV}\sim0.5$~mT, at point S.~\cite{Laird06} } 
\end{figure}

Pump-probe measurements were performed at nine  measurement points, $\epsilon_{\rm{M}}$.
Parametric plots of extracted visibility as a function of $\Delta B_z$  are shown in Fig.~\ref{T1Fig}(a) for three of the nine values. Note that for smaller values of $\epsilon_{\rm{M}}$, visibility decreases more rapidly with increasing $\Delta B_z$.

To separate the various contributions to visibility, we first define a total visibility, $V_{\rm{T}}$, as the product of $V_{\rm M}$, the extrinsic visibility of single-shot readout, whose contributions, such as amplifier noise, are discussed in Ref.~\onlinecite{Singleshotpaper}, and $V_J$, the intrinsic visibility of $S$-$T_{0}$ precession, reduced by a nonvanishing $J$ at the separation point  [Eq.~(\ref{CL})],
\begin{equation}
\label{visibility}
V_{\rm T}= V_{\rm M}\,V_J, 
\end{equation}
noting that $V_{J}\sim1$ for all but the smallest Overhauser gradients, $\Delta B_z\lesssim 1$~mT. $V_{\rm{M}}$ is calculated from the experimental parameters following Ref.~\onlinecite{Singleshotpaper}, and depends on the triplet relaxation time, $T_1$, at the measurement point. $T_{1}(\Delta B_z)$ is measured  via the charge signal, $v_{\rm{rf}}$, with 100 ns time resolution over $\tau_{\rm{M}}^{\rm{max}}= 4~\mu$s. After 200 s, the measurement is repeated with 250 ns time resolution over $15~\mu$s, allowing both short and long $T_{1}$ regimes to optimally fill the oscilloscope memory. Figure~\ref{T1Fig}(b) shows $\langle v_{\rm{rf}} \rangle$ averaged over 8000 probe cycles over a range of $\tau_{\rm{S}}$ values from 1 to 100~ns, as a function of measurement time, $\tau_{\rm{M}}$. An exponential fit to the decay of $\langle v_{\rm{rf}}(\tau_{\rm{M}}) \rangle$ from the short-time value $v_{\rm{rf}}^{(S)}+\Delta v_{\rm{rf}}P_T(0)$---corresponding to the initial mixture of charge states (0,2) and (1,1)---to the saturating value  $v_{\rm{rf}}^{(S)}$ yields relaxation times $T_1\sim 13~\mu$s for $\Delta B_z=3$~mT and  $T_1\sim 0.8~\mu$s for $\Delta B_z=15$~mT.  

Similar to the visibility, $T_1$ decreases  with increasing $\Delta B_z$ [Fig.~4(c)] and  decreasing $\epsilon_{\rm{M}}$ [Fig.~4(d)]. To compare the model to the experimental data, the value of $J(\epsilon_{\rm{M}})$ is determined by Eq.~(\ref{Jofepseqn}) with the triplet tunnel coupling, $t_T\sim12~\mu$eV as a single fit parameter. The singlet tunnel coupling, $t_S\sim10~\mu$eV, is estimated from the detuning, $\epsilon_{\rm{I}}$, of the $S$-$T_+$ resonance.~\cite{tsubs} The energy detuning, $\epsilon_T\sim300~\mu$eV, of the triplet charge transition is determined from dc transport measurements.~\cite{JohnsonSpinBlockade,VanderWiel} The fit, together with the measured parameters, yields the $\epsilon$-dependence of exchange energy, $J(\epsilon_{\rm{M}})$, shown in Fig.~\ref{levels}(c). 

The bare triplet relaxation rate, $\Gamma_T\sim(40~\mu{\rm{s}})^{-1}$, is assumed to be equal for all detunings, an approximation that is justified by the weak dependence of Eq.~(\ref{T1eqn}) on $\Gamma_T$ for $\Delta B_z>1$~mT and because measured values of $T_1$ agree with each other within the errors at small $\Delta B_z$. For the singlet charge relaxation rate, $\Gamma_S(\epsilon_{\rm{M}})$, one fit parameter is used for each detuning, $\epsilon_{\rm{M}}$, yielding the values shown in Fig.~\ref{T1Fig}(e). A fit of  $\Gamma_S(\epsilon_{\rm{M}})$ to the form Eq.~(\ref{GammaSeqn}) gives $\alpha\sim 11~\mu{\rm{eV}}\,{\rm{ns}}^{-1}$ and $\beta\sim 1600~\mu{\rm{eV^2\,ns}}^{-1}$. At $\epsilon_{\rm{M}} \sim 150~\mu$eV, the contributions from 2D and 3D phonons are about equal, and the fit is in reasonably good agreement with the data. The charge relaxation rates are consistent with the values measured in Ref.~\onlinecite{Fujisawa98}, when taking into account the difference in tunnel couplings, $t_S$. Deviations from the form (\ref{GammaSeqn}) are expected, e.g. due to resonances from finite lengths in the phonon environment.~\cite{Fujisawa98} 

Figure~\ref{T1Fig}(d) shows the model, Eq.~(\ref{T1eqn}), with $\Gamma_S(\epsilon_{\rm{M}})$ from Eq.~(\ref{GammaSeqn}), using $\alpha$ and $\beta$ from the fit in Fig.~\ref{T1Fig}(e). Extracted values of $\alpha$ and $\beta$ are rough estimates, as the functional form of $J(\epsilon_{\rm{M}})$, Eq.~(\ref{Jofepseqn}), is only approximate. The detuning dependence of $\Gamma_S$, assuming Eq.~\ref{GammaSeqn} and using the obtained fit parameters $\alpha$ and $\beta$, is shown in Fig.~\ref{levels}(c). Since $\Gamma_S \propto t_S^2$,~\cite{Fujisawa98} and roughly $J\propto t_S^2$  ($t_T$ increases with $t_S$), the first and dominant term in Eq.~(\ref{T1eqn}) becomes $\propto \Delta B_z^2/t_S^2$ for $\Delta B_z<J$. Contrary to intuition, a more transparent tunnel barrier yields longer triplet relaxation times, which is beneficial for quantum information processing, where large tunnel couplings enable fast operations.~\cite{petta05,Foletti09} 

\begin{figure}[t]
\includegraphics{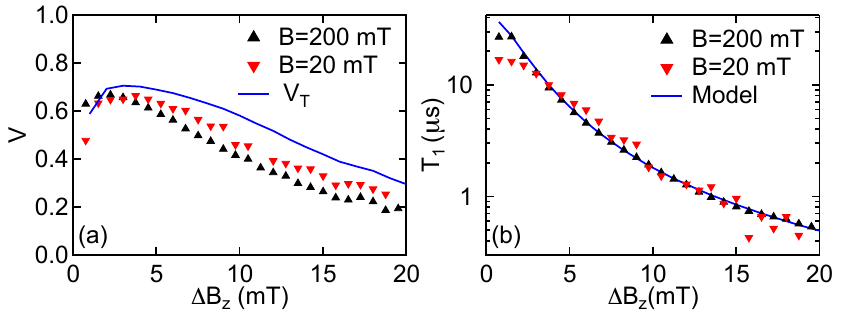}
\caption{\label{FieldFig} (Color online) (a) Parametric plot of measured visibility, $V$ [Fig.~3(f)], versus nuclear field difference, $\Delta B_z$[Fig.~4(e)] for applied fields of 200 mT and 20 mT.  Model of total visibility, $V_{\rm{T}}$, using  measured relaxation time, $T_1$ [Fig.~3(f)]. (b) Triplet relaxation time, $T_1$, as function of nuclear field difference, $\Delta B_z$, along with the model, using Eq.~(\ref{T1eqn}). The $\Delta B_z$ dependence of $V$ and $T_1$ does not depend on the applied magnetic fields.
} 
\end{figure}
The magnitude of the applied magnetic fields changes the overall rate of decay of nuclear polarizations [Figs.~\ref{timeev}(e,f)]. However, the parametric dependence of visibility and relaxation time on Overhauser gradient do not change with applied magnetic field, $B$, as shown in Fig.~5. 

\begin{figure}
\includegraphics{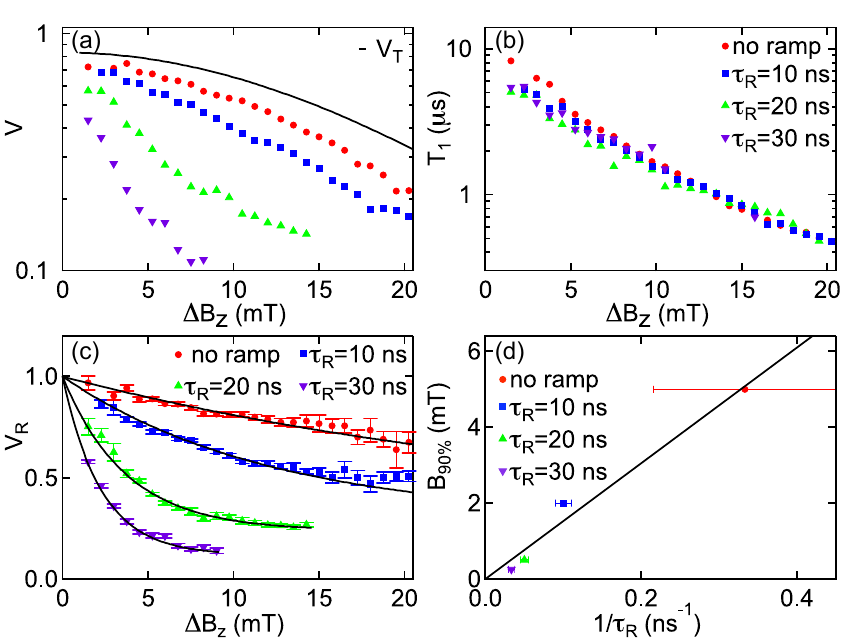}
\caption{\label{RampFig} (Color online) (a)  Parametric plot of visibility, $V$, of $S$-$T_0$ precession as a function of field gradient, $\Delta B_z$, for ramp times $\tau_{\rm{R}}$ [legend in (b) and (c)] from initial singlet to separation point S and back to measurement point M, along with model total visibility, $V_{\rm{T}}$ (solid curve) based on measured $T_1$ values. (b) Relaxation time, $T_1$, as function of nuclear field difference, $\Delta B_z$, shows no dependence on ramp time, $\tau_{\rm{R}}$. (c)  Ramp-rate visibility factor $V_{\rm{R}} = V / V_T$ as function of $\Delta B_z$ for different $\tau_{\rm{R}}$, along with exponential fits, $V_{\rm{R}} = e^{-\Delta B_z/B_{\rm{W}}}+V_0$~\cite{Fig6Params} (d) $B_{90\%}$, the nuclear field difference for which $V_{\rm{R}}=0.9$, as function of inverse ramp duration, $1/\tau_{\rm{R}}$. Without an intentional ramp, $ \tau_{\rm{R}}\sim3~{\rm{ns}}$. 
} 
\end{figure}

Finally, we consider the effect of finite pulse rise times on visibility. We observe that the model for total visibility, $V_{\rm{T}}$, based on Eq.~(\ref{visibility}), using measured values of $T_1$, overestimates the measured visibility except at the lowest values of $\Delta B_z$ [Fig.~5(a)]. We attribute this deviation to finite ramp rates of the separation and return pulses, which we can account for phenomenologically by including a ramp-rate-dependent factor $V_{\rm{R}}$ in the visibility model,
\begin{equation}
\label{measuredV}
V =  V_{\rm{R}}\, V_{\rm{T}}.
\end{equation}
 We investigated  $V_{\rm{R}}$ by including a deliberate ramp time, $\tau_{\rm{R}}$, to the probe cycle.  Following a 60 s, 4~MHz pump-cycle sequence, a prepared singlet was separated to point S over the ramp time, $\tau_{\rm{R}}$, then ramped to point M, also over $\tau_{\rm{R}}$. Resulting visibilities are shown in Fig.~\ref{RampFig}(a), together with the model $V_{\rm{T}}$ based on the measured $T_1$ using Eq.~(\ref{visibility}). Note that $T_1$ itself does not depend on $\tau_{\rm{R}}$, as shown in Fig.~6(b). The ramp-rate factor, $V_{\rm{R}}$, extracted by dividing the measured visibility by the model total visibility, $V_{\rm{R}}=V/V_{\rm{T}}$, is  shown in Fig.~6(c). With longer ramp times, $V_{\rm{R}}$ decreases more rapidly with increasing $\Delta B_z$. Due to the finite rise time of the pulses, the data without intentional ramp time has an estimated ramp time, $\tau_{\rm{R}}\sim3$~ns, and  $V_{\rm{R}}<1$, at finite nuclear field differences. 
 
 To further characterize the dependence $V_{\rm{R}}$ on $\Delta B_z$, we define $B_{90\%}$ as the maximum gradient for which $V_{\rm{R}}>0.9$. A phenomenological exponential fit [Fig.~6(c)] gave values for $B_{90\%}$ that increase roughly linearly with increasing ramp rate, $1/\tau_{\rm{R}}$ [Fig.~6(d)]. We conclude that the visibility factor $V_{\rm{R}}$ for the probe-cycle without an intentional ramp can similarly be attributed to the finite pulse rise time, limited by the bandwidth of the experimental set-up.

\section{Conclusions}
\label{conclusions}
An enhanced Overhauser field gradient results from an electron-nuclear spin pumping cycle under all conditions investigated. The resulting difference in $z$ components of Overhauser fields reduces the relaxation time of the $m=0$ triplet state during measurement, lowering the visibility of single-triplet qubit readout. Bandwidth limited pulses further reduce readout fidelity.  Visibility reduction due to field gradients appears to be the likely explanation for the experiments discussed in Ref.~\onlinecite{ReillyZamboni}. For applications using magnetic or Overhauser field gradients~\cite{Foletti09,PioroLadriere} it is desirable to design the exchange interaction to allow long triplet lifetimes.  In the presence of a magnetic field difference, the device should be tuned to a large inter-dot tunnel coupling with a measurement point chosen at large detuning, near the $T_{0}$ anticrossing, where exchange protects the triplet, while the charge relaxation rate is small. To mitigate errors from finite pulse rise times, an initialization of the qubit via adiabatic loading of $\mid \uparrow \downarrow \rangle$, followed by a $\pi/2$ pulse, may be preferable over the diabatic initialization used here and in Refs.~\onlinecite{petta05,Foletti09,ReillyCorr07}. We would like to point out that the results of this paper do not imply a short relaxation time of the qubit while it is operated in the (1, 1) state, where $T_1$ is much longer and expected to be independent of magnetic field gradients.~\cite{Amasha06} \\
\begin{acknowledgments}
We acknowledge funding from IARPA/MQCO program and DARPA/QUEST program. Device fabrication used Harvard's Center for
Nanoscale Systems (CNS), supported by the National
Science Foundation under ECS-0335765. We thank D.J. Reilly, J.M. Taylor, and S. Foletti for useful discussion.
\end{acknowledgments}



\begin{thebibliography}{14}
\bibitem{leverarmdisclaim} The leverarms for gate voltages $V_{\rm{L}}$ and $V_{\rm{R}}$ are approximately equal in the measured device.
\bibitem{Jsneglect}
$J_{\rm{S}}$, estimated to be $J_{\rm{S}}\sim10~{\rm{neV}}\pm5~{\rm{neV}}$, from the drop of $V_J$ at the lowest measured field differences,  results in a shifted frequency $f_{\rm{S}}^* = \sqrt{f_{\rm{S}}^2+(J_{\rm{S}}^2/h)^2}$,  a negligible correction at the relevant frequencies.
\bibitem{Param3b}
Other parameters are, $P_0= 0.6$ for $\Delta t = 150$ s, and $P_0= 0.4$ for $\Delta t = 300$ s.
\bibitem{Fig6Params}
 For no ramp, $V_0 \sim 0.3$, $B_{\rm{W}}\sim31$~mT; for $\tau_{\rm{R}}=10$~ns, $V_0 \sim 0.3$, $B_{\rm{W}}\sim12$~mT; for $\tau_{\rm{R}}=20$~ns, $V_0 \sim 0.2$, $B_{\rm{W}}\sim4$~mT; for $\tau_{\rm{R}}=30$~ns, $V_0 \sim 0.1$, $B_{\rm{W}}\sim2$~mT. In Fig.~\ref{RampFig}(d), the slope of the linear fit is  $\sim 15~\rm{mT\,ns}$.
\end{thebibliography}
$^*$These authors contributed equally to this work.\\
$^{\dagger}$ Present address: 2nd Institute of Physics C, RWTH Aachen University, DE-52074 Aachen, Germany

\end{document}